# Physics the Google Way

**David W. Ward,** Massachusetts Institute of Technology, Cambridge, MA

*"Never memorize something that you can look up."*—Albert Einstein

Are we smarter now than Socrates was in his time? Society as a whole certainly enjoys a higher degree of education, but humans as a species probably don't get intrinsically smarter with time. Our knowledge base, however, continues to grow at an unprecedented rate, so how then do we keep up? The printing press was one of the earliest technological advances that expanded our memory and made possible our present intellectual capacity. We are now faced with a new technological advance of the same magnitude—the internet—but how do we use it effectively? A new tool is available on Google (http://www.google.com) that allows a user not only to numerically evaluate equations, but to automatically perform unit analysis and conversion as well, with most of the fundamental physical constants built in.

## The Fundamental Constants

To get a feel for how the Google calculator works, you can start by googling[1] "`2+2=`". You can click on "`More about calculator,`" located directly below the result, to learn the basics of Google equation formatting and there you will find a link to more complete instructions on how to format equations for Google, e.g. "`x^n`" means raise x to the n$^{th}$ power, "`sqrt(x)`" means take the square root of x, etc. Return to the search bar and google "`G`", as illustrated in figure 1. Google returns "`gravitational constant = 6.67300 × 10⁻¹¹ m³ kg⁻¹ s⁻²`" at the top of the page next to the calculator icon, followed by web page results for the Google query. The results relevant to this article will always be those at the top of the page adjacent to the calculator icon; only these can be used in equations. Google has most of the fundamental constants built in, many of which are listed in table 1.[2]

To illustrate the use of the constants in an equation, consider the fine structure constant, in which the fundamental constants of quantum mechanics, electricity and magnetism, and geometry are contained. We will actually google one over the fine structure constant, as this has a well know simple value. Google "`(4 * pi * electric constant * hbar * c) / (elementary charge^2) =`" to find that it is equal to "`137.035984`". We could have just googled "`1 / fine-structure constant`" to find the same answer. Note, we were able to confirm that our calculation not only matches the acceptable numerical value of the fine structure constant, but is unitless as well.

## Unit Conversions

The previous example illustrates the utility of the Google calculator in unit checking, a task which otherwise is only adequately described by the word tedium. Although SI units have solid footing as a scientific standard, it is sometimes useful to employ others, or variants of SI units. This is not only because many classic texts and some individual fields in physics employ units other than SI, but mainly because these are not necessarily the units in which we understand the world. In the United States, a comfortable room

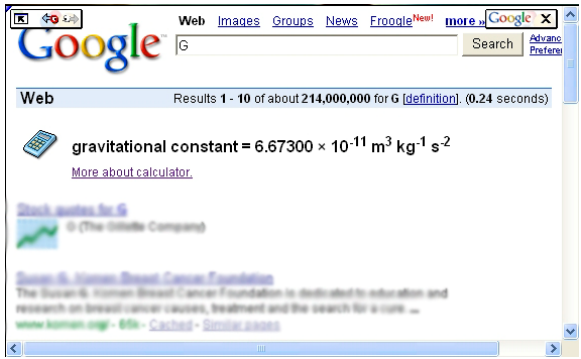

**Fig. 1.** Type equations directly into the search bar, and if applicable, Google outputs the result adjacent to a calculator icon. Click on "More about calculator" to learn Google formatting of equations.

is one that is cooled to around 70 degrees Fahrenheit, not 21.1 degrees Celsius or 294.3 Kelvin.

## Illustrative Examples

To further illustrate the advanced features of the Google calculator, we will consider several example calculations, exploiting the Google features to tailor the form of the answer we obtain.

### How much do I weigh on the moon?—the sun?

Space travel is fun, and though the internet won't allow us to do it from the comfort of our homes, we can get a sense of what it would be like to visit another planet, without the harmful side effects of toxic atmospheres and unbearable temperatures, see figure 2. Since our weight, not our mass, is just the gravitational attraction we experience to the local gravity, we can use Newton's law of gravity to determine what we would weigh on other planets. If on earth our weight was 100 pounds of force, which is the number our bathroom scale indicates, then our mass would be 100 pounds. If we took our scale to the moon then to find out what the scale would read there, google "`convert (G*mass of the moon*100 pounds/(radius of the moon)^2) to pounds force`" to find the answer "`16.5912317 pounds force.`" Replace `the moon` with `the sun` to find that we would need to take a better scale as our weight there would be "`2,797.844 pounds force.`" Try it for Mars or any other planet in the solar system by using "`mass of planet name`" and "`radius of planet name`" in Newton's law of gravity.

### Einstein's Car

Let's suppose you built a crazy car that converted mass into energy directly. So you put in a pound of matter, perhaps old cereal boxes, and the engine of this car converts it directly into energy through Einstein's famous energy/matter relation[3], $E=mc^2$—maybe we have to put in an equal dose of anti-matter to get 100% conversion, but we'll assume that somehow it does. Let's suppose that it is capable of converting one pound per year under constant operation. How many horsepower does this engine have? Google "`convert 1 pound* c^2 / year to horsepower`" and it returns "`1,732,400.85 horsepower.`" It makes a Ferrari look like a go-cart.

### Internal Temperature of the Sun

Inside the sun is a fusion engine creating heat and pressure that push everything outward from the center, but there is also gravity pulling everything in towards the center. Since the sun does not appear to be expanding or compressing, the pressure from the heat engine at the center must be equal to the gravitational pressure. If we assume that the sun is composed entirely of hydrogen, i.e. protons, and that the hydrogen atoms behave like an ideal gas within the volume of the sun, then the temperature due to this internal pressure is given by the familiar ideal gas law, $PV=nKT$, where $P$ is the internal pressure, $V$ is the volume of the sun, $n$ is the number of hydrogen atoms (protons) that make up the sun, $K$ is Boltzmann's constant relating heat energy to temperature, and $T$ is the temperature of the sun, which is the quantity we seek. The pressure from the sun's gravity is just the force of the sun's gravity on itself divided by the surface area of the sun. Plugging all of this into the ideal gas law and solving for $T$, we can google the resulting equation to find our estimate of the Sun's temperature at its center: "`(G * proton mass * mass of the sun) / (Boltzmann constant *` radius of the sun) = 23,118,268.8 Kelvin, or ~$2.3 \times 10^7$ K." Compared to the accepted value of $1.5 \times 10^7$ K, this is not a bad estimate considering the approximations used. By the way, how did I know to use the Boltzmann rather the Rydberg constant in this problem? I originally used the Rydberg constant, but got an answer in Kelvins/mol, while I expected an answer in Kelvins.

*Electrostatic Versus Gravitational Force*

Physics as we know it was born with Sir Isaac Newton's *Mathematical Principles of Natural Philosophy* or simply *the Principia* in 1687.[4] The seminal equation from this work is the law of gravitational attraction, which contains the gravitational constant, *G*. To illustrate the use of the gravitational law, google "`G*mass of the earth*mass of the sun/(1 astronomical unit)^2`" to find that the force binding the earth to the sun is "`3.54296305 × 10`$^{22}$` Newtons`." Using the equation for centripetal acceleration, $a=v^2/r$, the earth's rotational speed can be determined. To do so google "`sqrt(G*mass of the sun/(1 astronomical unit))`" to learn that the earth is hurling through space at the rate "`29,785.5982 m/s`."

One hundred years after the publication of the *Principia*, Charles Augustin Coulomb published, in his second memoir on electricity and magnetism, the law that bares his name; Coulomb's law is analogous to Newton's law of gravitation, with the masses replaced by charge, and the gravitational constant replaced by the Coulomb constant which is now known to be $1/4\pi\varepsilon_0$, where $\varepsilon_0$ is the permittivity of free space, or electric constant. How does the electric force between the electron and proton in the hydrogen atom compare to the gravitational force between them? To find the ratio of the electric to gravitational force in the hydrogen atom google "`((elementary charge)^2 / (4 * pi*electric constant))/(G*mass of electron*mass of proton)`" to discover that the electric force is "`2.26910384 × 10`$^{39}$" times stronger.

| long Name | Shorthand |
|---|---|
| atomic mass units | amu or u |
| Astronomical Unit | au |
| Avogadro's number | N_A |
| Bolzmann constant | k |
| electric constant, permittivity of free space | epsilon_0 |
| electron mass | m_e |
| electron volt | eV |
| elementary charge | |
| Faraday constant | |
| fine-structure constant | |
| gravitational constant | G |
| magnetic constant, permeability of free space | mu_0 |
| magnetic flux quantum | |
| mass of the moon | m_moon |
| mass of the sun | m_sun |
| mass of **[planet name]** | m_mercury,… |
| molar gas constant | R |
| Planck's constant | h and hbar |
| proton mass | m_p |
| radius of the moon | r_moon |
| radius of the sun | r_sun |
| radius of **[planet name]** | r_mercury,… |
| Rydberg constant | |
| speed of light | c |
| speed of sound (note: in air at sea level) | |
| Stefan Boltzmann constant | |

**Table 1. Guide to physical constants available on Google calculator. Entries are sometimes case sensitive.**

Had we simply asked for the electrical force between them instead of the ratio we would find that the force is equal to "`1.85205315×10`$^{-08}$` pounds force`," using the Bohr radius as the separation distance between them.[5] What can we think of that has a comparable weight in the earth's gravity? Approximate a human cell as a sphere of radius ten microns (kind of a big cell) composed entirely of water. How much does this sphere weigh in earth's gravity? Google "`convert 1000 kg/liter *4*pi*(10 microns)^3 to pounds`" to find "`2.77041049 × 10`$^{-08}$` pounds`," which is the same amount in pounds of force in earth's gravity. So the electrical force between the proton and electron is roughly the same as the

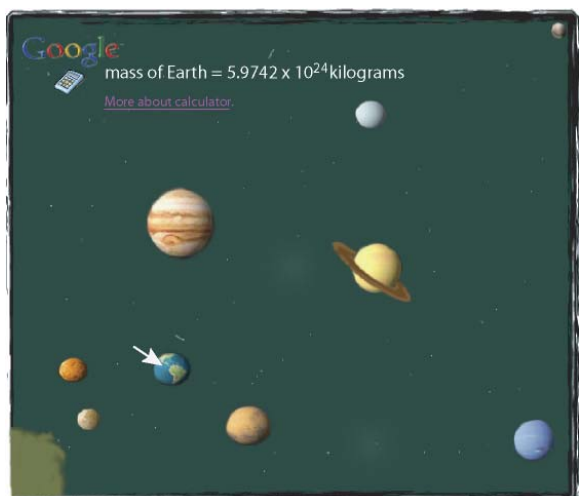
weight of a human cell in the earth's gravity.

**Figure 2: The Google calculator has planetary data about our solar system built in.**

However, the electron has a much smaller mass, and this force creates a ridiculously large acceleration on the electron. Using the Bohr model of the hydrogen atom, we can compute the rotational kinetic energy of the electron were it orbiting the nucleus as in a planetary system by googleing "`convert 0.5 * ((elementary charge)^2 / (4*pi*electric constant *(0.529189379 angstrom))) to eV.`" The answer is "`13.6053788 electron volts,`" which is darn close to the measured value for the ionization energy of hydrogen. We could have just obtained the answer in Joules, but no one thinks about atomic energy levels in Joules.

## Conclusion

I began this article with a quote from Albert Einstein. Given the vast quantity of information available on any given subject in physics, this makes a lot of sense, but we all know how useful it is to memorize certain things before an exam, for example. The point is that there is advantage at having information at your fingertips. The advanced features of the Google calculator make this possible without us having to do the memorizing.

Future generations of physicists will pass the responsibility of unit conversion, unit checking, algebra and calculus, and looking up the physical constants to computers much in the same way our generation passed off addition, subtraction, multiplication, and division. The earlier this responsibility is passed off in a student's education, the sooner they can get to forefront of physics. They will, of course, need some training in the basics, but we will eventually abandon spending years on multiplication tables, algebra, and calculus so that students can tackle introduction to string theory before they graduate from high school. They won't be able to do it because they got intrinsically smarter, but because the Google calculator bar in their heads-up-display which is linked to the internet through their personal WI-FI connection will enable them to concentrate on the really important things without spending a lot of time worrying about the remedial math.

**David W. Ward** received his B.S. in physics from the College of Charleston and is presently completing his Ph.D. at the Massachusetts Institute of Technology. His current research is in novel optical materials, nonlinear optics, and computational electrodynamics.
**Department of Chemistry, Massachusetts Institute of Technology, Cambridge, MA 02139; david@davidward.org**